\documentclass[12pt,letter]{article}

\usepackage{amsmath}
\usepackage{amsfonts}
\usepackage{dsfont}
\usepackage{amsxtra}
\usepackage{amssymb}

\usepackage{graphicx,epsfig}

\newcounter{theappend}
\newcommand{\append}[1]{
\addtocounter{theappend}{1} 
\addcontentsline{toc}{section}{\protect \numberline{\Alph{theappend}.}{{\rm\bf #1}}}
\pagebreak[3] \par\vskip 1.5cm
\noindent{\large\bf \Alph{theappend}.\;\; #1}
\nopagebreak[4]\par\vskip .0cm\noindent}

\newcounter{apps}

\newcounter{prs}[section]

\newcounter{cors}

\newcounter{figs}

\newcommand{\be}{\begin{equation}}
\newcommand{\ee}{\end{equation}}
\newcommand{\bea}{\begin{eqnarray}}
\newcommand{\eea}{\end{eqnarray}}
\newcommand{\beaa}{\begin{eqnarray}}
\newcommand{\eeaa}{\end{eqnarray}}

\newcommand{\ba}{\begin{array}}
\newcommand{\ea}{\end{array}}

\newcommand{\lb}{\label}

\newcommand{\ra}{\rightarrow}
\newcommand{\lra}{\longrightarrow}

\newcommand{\td}{\tilde}
\newcommand{\e}{\epsilon}
\newcommand{\al}{\alpha}

\newcommand{\p}{\partial}

\newcommand{\ld}{\lambda}
\newcommand{\Ld}{\Lambda}
\newcommand{\vp}{\varphi}

\newcommand{\Om}{\Omega}

\newcommand{\Tr}{{\rm Tr}}

\newcommand{\N}{{\mathcal{N}}}

\newcommand{\Z}{{\mathbb{Z}}}

\newcommand{\C}{{\mathbb{C}}}

\newcommand{\PP}{{\mathbb{C}}{\rm P}}

\begin{document}

\begin{titlepage}

\begin{flushright}
PUPT-2245\\
arXiv:0710.5519
\end{flushright}

\vspace{2cm}

\begin{center}

{\Large\bf On the Geometry of Metastable  \\[3mm] Supersymmetry Breaking}\\[1.5cm]
\bigskip
\bigskip
{Matthew Buican, Dmitry Malyshev%
\footnote{On leave from ITEP, Russia, Moscow, B. Cheremushkinskaya, 25},
 and Herman Verlinde} \\[1cm]

 \it { Physics Department,
 Princeton University,\\[1.5mm]
 Joseph Henry Laboratories, Princeton NJ 08544}\\[5mm]

\bigskip
{\bf Abstract}

\end{center}
We give a concise geometric recipe for constructing D-brane gauge theories that exhibit
metastable SUSY breaking. We present two simple examples in terms of branes at
deformed CY singularities.

\end{titlepage}

\newpage
\tableofcontents

\section{Introduction}

\noindent
Finding robust string realizations of gauge theory models that
exhibit dynamical SUSY breaking (DSB) is an important facet of
string phenomenology. While recent studies have uncovered a growing
number of string systems with DSB, there are only a few examples known
in which the SUSY breaking mechanism is well understood from both
the gauge theory side and the geometric string perspective.

The dynamical mechanism of SUSY breaking assumes that the lagrangian
is supersymmetric but that, due to non-perturbative dynamics,
the vacuum configuration breaks SUSY at an exponentially low scale \cite{DSB}.
In general, such non-supersymmetric vacuum
states need not be the true vacuum of the theory, but may instead
represent long-lived metastable states. While controlled examples of
metastable vacua in  string theory have been known for some time
\cite{KPV}, the increased recent interest in their manifestations
and properties was sparked by the discovery by Intriligator,
Seiberg, and Shih  of a metastable SUSY breaking vacuum for SQCD
with $N_f\! > \! N_c$ massive flavors \cite{ISS}. Realizations of
the ISS mechanism in string theory, as well as other stringy systems
with metastable SUSY breaking, have since been found \cite{KachruDSB}%
\cite{Ooguri:2006bg}\cite{Aganagic:2006ex}\cite{Douglas:2007tu}.
Another recent advancement has been to merge the calculational power
of geometric transitions
 with insights from field theory
to engineer basic field theoretic models of SUSY breaking
\cite{Aganagic:2007py}.

%

In this paper, we will consider gauge theories on D-branes near a
singularity inside a Calabi-Yau manifold. Our goal is to identify a
general geometric criterion for the existence of F-type SUSY
breaking, and to use this insight to construct simple examples of
D-brane systems that exhibit metastable SUSY breaking. F-type SUSY
breaking corresponds to the unsolvability of F-term equations
\bea
\frac{\p W}{\p \Phi}\neq 0
\eea
where $W(\Phi)$ is a superpotential depending on the chiral field
$\Phi$. The simplest example of this type is the Polonyi model,
consisting of a single chiral field with superpotential $W(\Phi)
=f\Phi$ in which SUSY is broken by the non zero vacuum energy $V
\sim |f|^2$.

We will assume that the non-perturbative dynamics manifests itself
in deformations of a theory with unbroken SUSY.
The main purpose of our paper is to study the consequences of these deformations.
In the context of D-branes in IIB string
theory, deformations of the superpotential correspond to complex
deformations in the local geometry. The deformed geometry still
satisfies the Calabi-Yau condition and  the D-brane lagrangian is
fully supersymmetric but the vacuum configuration of
the gauge theory breaks SUSY spontaneously. In the geometric
setting, this corresponds to a D-brane configuration that, while
submerged inside a supersymmetric background, gets trapped in a
non-supersymmetric ground state.

As a simple illustrative example, consider type IIB string theory on
a $\C^2/\Z_2$ orbifold singularity
\cite{Douglas:1996sw}\cite{Klebanov:1998hh}, with $N$ fractional
D5-branes wrapped on the collapsed 2-cycle. The corresponding field
theory consists of a $U(N)$ gauge theory with a complex adjoint
chiral field $\Phi$. Since the $\Z_2$ orbifold locus defines a
non-isolated singularity inside $\C^3$,
 the fractional D5-branes are free to move
along a complex line. The location of the $N$ branes along the
non-isolated singularity is parameterized by the $N$ diagonal
entries of the complex field $\Phi$. As we discuss in more detail in
section 2, there exist a deformation of the singularity that
corresponds to adding the F-term
\bea
W=\zeta\, \Tr\Phi
\eea
to the superpotential. Geometrically, the parameter $\zeta$ is
proportional to the period of the holomorphic two-form over the
deformed 2-cycle. The fractional D-brane gauge theory then breaks
SUSY in a similar way to the Polonyi model. This simple observation
lies at the heart of many type IIB D-brane constructions of gauge theories that
exhibit F-term SUSY
breaking.\footnote{F-term SUSY breaking in type IIB
D-brane constructions naturally involves deformed non-isolated
singularities, that support finite size 2-cycles which D5-branes can
wrap  \cite{Douglas:1996sw}. Deformations of isolated
singularities correspond to
3-cycles that in type IIB cannot by wrapped by the space-time filling D-branes.\\[-4mm]}
SUSY breaking via D-terms can be described analogously.\footnote{
Turning on the FI parameters of the type IIB D-brane gauge theory amounts to
blowing up the collapsed two-cycles of a CY singularity. These
blowup modes are K\"ahler deformations of the geometry, and are
somewhat harder to control in a type IIB setup
than  the complex structure deformations that we use in our study. 
}

We wish to use this simple geometric insight to construct more
interesting gauge theories with DSB, and in particular, with
ISS-type SUSY breaking and restoration. When viewed as a quiver
theory,  the ISS model has two nodes, a ``color'' node with gauge
group $SU(N)$ and a ``flavor'' node with $SU(N_f)$ symmetry. The
``flavor" node has an adjoint field. This suggests that the flavor
node must be represented by a stack of $N_f$ fractional branes on a
non-isolated singularity. The ``color" node, on the other hand, does
not have an adjoint, and thus corresponds to branes that are bound
to a fixed location. The natural representation for the color node
is via  a stack of $N$ branes placed at an isolated
singularity.

Our geometric recipe for realizing an ISS model in IIB string theory
is as follows:

\bigskip

\noindent ${}$ \ \parbox{15.9cm}{\addtolength{\baselineskip}{0mm}
\addtolength{\parskip}{1mm}

$1.$ Find a Calabi-Yau geometry with a non-isolated singularity
passing through an isolated\\ ${}$\, \; singularity such that there
exists a deformation of the non-isolated singularity.

\noindent $2.$ Put some number of D-branes on the isolated
singularity and some number of fractional\\ ${}$\, \; branes on the
non-isolated singularity.\, By conservation of charge, the
branes can not\\ ${}$\, \; leave the non-isolated singularity.

\noindent 3. When we deform the non-isolated singularity, an F-term
gets generated that results in \\ ${}$\; \; dynamical SUSY
breaking. \, The fractional branes have a non-zero volume, and their
\\ ${}$\, \; tension lifts the vacuum energy above that of the SUSY
vacuum.

 \noindent
4. There is a classical modulus corresponding to the motion of the
fractional branes along\\ ${}$\; \; the non-isolated singularity.
This modulus can be fixed in a way similar to ISS, by the\\ ${}$\;
\; interaction with the branes at the isolated singularity.}
\bigskip

Following this recipe we will geometrically engineer, via an
appropriate choice of the geometry and fractional branes,
 gauge theories that are known to exhibit meta-stable DSB.
The eventual goal is to fully explain in
geometric terms all field theoretic ingredients: the field content
and couplings, the meta-stability of the SUSY-breaking vacuum,
and the process of SUSY restoration. While in our examples we will
be able identify all these ingredients, we will not have sufficient dynamical
control over the D-brane set-up to in fact {\it proof} the existence
of a meta-stable state on the geometric side. Rather,
by controlling the geometric engineering dictionary,
we can rely on the field theory analysis to demonstrate that the
system has the required properties.

This wish to have geometrical control over the field theory parameters
also motivates why we prefer to work with local IIB D-brane constructions.
Although we will work in a probe approximation, in principle we could
extend our analysis to the case where the number of branes becomes large.
In  this AdS/CFT limit, there should exist a precise dictionary between the
couplings in the field theory and the asymptotic boundary conditions
on the supergravity fields \cite{ads/cft}.
By changing these boundary conditions one can tune the UV couplings.
This in principle allows full control over the IR couplings and dynamics.

The organization of the paper is as follows. In section 2, as a
warm-up, we discuss the F-term deformation of D-branes on
$\C^2/\Z_2$. In section 3 we describe the realization of meta-stable
supersymmetry breaking via D-branes on the suspended pinch point
singularity. We find that supersymmetry restoration involves a
geometric transition. In section 4 we give the IIA dual description
of the same system and find that it is similar to the IIA
constructions of \cite{Ooguri:2006bg, IIAfollow}.
Finally, in section 5, we present a  D-brane
realization of the Intriligator-Thomas-Izawa-Yanagida model
\cite{IT, IY}, as an example of a system in which the F-term, that
triggers SUSY breaking, is dynamically generated via a quantum
deformation of the moduli space.

\smallskip

When this paper was close to completion, an interesting paper
\cite{Aganagic:2007py} appeared in which closely related results
were reported.\footnote{The IIB string realizations of DSB found in
\cite{Aganagic:2007py} were motivated by the earlier related
work \cite{Aharony:2007db} in type IIA theory, and by
the idea of retrofitting simple systems with DSB, put forward in \cite{Dine:2006gm}.}
 In agreement with our observations, in
\cite{Aganagic:2007py} the F-term SUSY breaking takes place due to
the presence of fractional D5-branes on slightly deformed
non-isolated singularities. One of the main points in
\cite{Aganagic:2007py} was to
 show that the deformation
can be computed exactly in the framework of geometric transitions:
this is an important step in finding calculable examples of SUSY
breaking in string theory. The main point of our paper is to
identify simple geometric criteria for the existence of SUSY
breaking vacua that can have more direct applications in model
building.

\newpage


\section{Deformed $\C^2/\Z_2$}

\noindent
The $\C^2/\Z_2$ singularity, or $A_1$ singularity, is described by the following complex equation
in $\C^3$
\be
\qquad cd=a^2, \qquad \qquad \quad \mbox{\small $(a,b,c) \in \C^3\, .$}
\ee
A D3-brane on $\C^2/\Z_2$  has a single image brane.
The brane and image brane recombine in two fractional branes. Correspondingly,
the quiver gauge theory for $N$ D3-branes at the $A_1$
singularity has two $U(N)$ gauge groups. It also has two adjoint matter fields
$\Phi_1$ and $\Phi_2$ (one for each gauge group),
and two pairs of chiral fields $A_i$ and $B_j$ $i,j=1,2$ in the
bifundamental representations $(N,\bar N)$ and $(\bar N, N)$
\cite{Douglas:1996sw}\cite{Klebanov:1998hh}.
The superpotential reads
\bea
W=g\, \Tr\Phi_1(A_1B_2-B_1A_2)+g\, \Tr\Phi_2(A_2B_1-B_2A_1)
\eea
A D3-brane has 3 transverse complex dimensions. The transverse space
$\C^2/\Z_2 \times \C $ has a non-isolated $A_1$ singularity.  It is therefore
possible to separate the fractional branes.
This corresponds to giving different vevs to the two adjoint fields.
In the limit of infinite separation one can consider a theory with
only one type of fractional brane.
This theory consists of a $U(N)$ gauge field with one adjoint matter
field and no fundamental matter.

Let us add an $F$ and a $D$-term
\bea
W_F=\zeta\, \Tr(\Phi_2-\Phi_1),\qquad \qquad
V_D=\xi\, \Tr(D_2-D_1).
\eea
The resulting F and D-term equations read
{\small
\bea
\lb{FD}
 \Phi_1 A_1 \! - \! A_1 \Phi_2 =0, & & \qquad A_2 \Phi_1  \! - \! \Phi_2 A_2 =0 , \qquad\quad A_1B_2-B_1A_2=\zeta,
 \nonumber\\[-2mm]\\[-2mm]
 \Phi_1 B_1 \! -\! B_1 \Phi_2 =0, & & \qquad  B_ 2 \Phi_1 \! - \! \Phi_2 B_2 =0 ,
 \qquad |A_1|^2+|B_1|^2\! -|A_2|^2\! -|B_2|^2=\xi.\nonumber
\eea}
These equations allow for a supersymmetric  solution, provided we set the adjoint vevs to be
equal, $\Phi_1 = \Phi_2$.
For generic $\zeta$ and $\xi$, some of the $A$ and $B$ fields acquire
vevs and break the $U(N)\times U(N)$ symmetry to a diagonal $U(N)$.
This corresponds to joining the $2N$ fractional branes into $N$ D3-branes.
The space of solutions of the F and D-term equations is the space where
the D3-brane moves, which turns out to be a deformed $A_1$ singularity described
by the equation%
\footnote{The general deformations of orbifold singularities of $\C^2$ where
found by Kronheimer \cite{Kronheimer:1989zs} as some hyperkahler
quotients.
Douglas and Moore noticed \cite{Douglas:1996sw} that these hyperkahler
quotients are described by the F and D-term
equations for D-branes at the corresponding orbifold singularities.}
\be
cd = a (a-\zeta),
\ee
where $c=A_1A_2$, $d=B_1B_2$ and $a=A_1B_2=B_1A_2+\zeta$ are the
gauge invariant combinations of the fields (in the last definition
we used the F-term equation for the $\Phi$ field).

The F-term coefficient $\zeta$ deforms the singularity,
the D-term coefficient, or FI parameter, $\xi$ represents
a resolution of the $\Z_2$ singularity.
In two complex dimensions both the resolution
and the deformation correspond to inserting a two-cycle,
$E\sim\PP^1$, instead of the singular point.
The parameters $\xi$ and $\zeta$ are identified with the periods of
the Kahler form and the holomorphic two-form on the blown up 2-cycle $E$
\bea
\label{period}
\xi=\int_E J\, ,\qquad \qquad \ \ \
\zeta=\int_E \Om^{(2)}\, .
\eea

The non-supersymmetric vacuum state arises in the regime where the
vevs of the two adjoint fields $\Phi_1$ and $\Phi_2$ are both different. Geometrically, this
amounts to separating the two stacks of fractional branes.
The bifundamental fields $(A_i,B_i)$, which arise as the ground states of open strings
that stretch between the two fractional branes, then become massive.
In the deformed theory, the F-term equations can not be satisfied and SUSY is broken.
In the extreme case, where one of the two stacks of fractional branes
has been moved off to infinity, so that e.g.  $\langle \Phi_2\rangle \to \infty$, the system
reduces to the Polonyi model: a single $U(N)$ gauge theory with a complex adjoint $\Phi_1$ and superpotential
$W= \zeta\, \Tr\Phi_1$.
The vacuum energy $V=N|\zeta|^2$ is interpreted as the tension of the $N$ fractional
branes wrapped over the deformed two-cycle.

Strictly speaking the single stack of fractional branes on a
deformed singularity is a supersymmetric configuration
(one manifestation is that the spectrum of particles in Polonyi
model is supersymmetric).
In order to break SUSY we really need the second stack of different
fractional branes on a large but finite distance.
In this case, the SUSY breaking vacuum is not stable due to the
attraction between the two stacks of branes.

\medskip

Before we get to our main example of the SPP singularity,
let us make a few comments:

1. The gauge theory on $N$ fractional branes on the $\C^2/Z_2$ singularity is an
$\N=2$ $U(N)$ theory.
If we deform the singularity, then SUSY is broken, whereas
in general, $\N=2$ theories are not assumed to have SUSY
breaking vacua (see, e.g., Appendix D of \cite{ISS}).
The point is that the SUSY breaking occurs in the $U(1)$ part of
$U(N)$ that decouples from $SU(N)$.
Moreover the $\N=2$ $U(1)$ theory consists of two non-interacting
$\N=1$ theories: a vector boson and a chiral field.
Thus the chiral field $\vp=\Tr\Phi$, responsible for SUSY
breaking, is decoupled from the rest of the fields in $\N=2$
$U(N)$ and SUSY is broken in the same way as in the Polonyi model.

2. In general, we consider $\N=1$ theories on isolated singularities
that intersect non-isolated singularities.
With appropriate tuning of the couplings,
the fractional branes wrapping the non-isolated cycles provide an
$\N=2$ subsector in the $\N=1$ quiver.
Removing the D-branes along the non-isolated singularity reduces the
field theory on their world volume to $\N=2$ SYM.
For this  reason the fractional branes on the non-isolated
singularity can be called $\N=2$ fractional branes
\cite{Franco:2005zu}\cite{Franco:2006es}.
Similarly to $\C^2/Z_2$ example, the presence of $\N=2$ fractional
branes on slightly deformed non-isolated singularity breaks SUSY.

3. The use of $\N=2$ fractional branes is the
distinguishing property of our construction from SUSY breaking
by obstructed geometry
\cite{Franco:2005zu}\cite{Berenstein:2005xa}\cite{Bertolini:2005di}.
The presence of the non-isolated singularity enables the relevant
RR-fluxes escape to infinity without creating a contradiction with
the geometric deformations.
In this way one can avoid the generic runaway behavior
(see, e.g., \cite{Intriligator:2005aw} \cite{Brini:2006ej}) of obstructed geometries
(in our case we still need to take the one loop corrections to the
potential into account in order to stabilize the flat direction along the
non-isolated singularity).


\section{ISS from the Suspended Pinch Point singularity}

\noindent
In this section, we will show how to engineer a gauge theory with ISS-type SUSY breaking
by placing fractional branes on the suspended pinch point (SPP) singularity.
First, however, we summarize the arguments that lead us to consider this
particular system.

As we have seen in the previous section,
several aspects of the ISS model are quite similar to the $\C^2/Z_2=A_1$ quiver theory.
The term linear in the adjoint in the
ISS superpotential is the $\zeta$ deformation of the $A_1$ singularity.
 Both models have two gauge groups
(the global flavor symmetry $SU(N_f)$ in ISS can be thought of as a
weakly coupled gauge symmetry).
The flavor gauge group is bigger than the color gauge group -- this can be achieved  in the $A_1$ quiver by introducing an excess of fractional branes of one type. The vevs of
bifundamental fields break
$SU(N_f)\times SU(N)\ra SU(N)_{diag}\times SU(N_f-N)$.
The breaking of $SU(N)\times SU(N)\ra SU(N)_{diag}$ corresponds to recombination
of $N$ pairs of fractional branes into $N$ (supersymmetric) D3-branes.
The vacuum energy is proportional to the tension of the remaining
$N_f-N$ fractional branes.

There is however an important difference between the two systems.
In ISS it is crucial that the color node $SU(N)$ doesn't have an
adjoint field and that all the  classical moduli are lifted by one loop
corrections.
In the $\C^2/\Z_2$ orbifold there is also an adjoint in the
\lq\lq flavor'' node.
Giving equal vevs to the two adjoints in the $\C^2/Z_2$ quiver
corresponds to the \lq\lq center of mass'' motion of the system of branes
along the non-isolated singularity.
This mode doesn't receive corrections and remains a flat direction.

Thus, the key distinguishing feature of ISS relative to the $\C^2/\Z_2$ model
is that the color gauge group $SU(N)$ has no adjoints.
For constructing a geometric set-up, we
need a mechanism that fixes the position of the $N$ D3-branes.
The gauge theories without adjoint fields  are naturally engineered by
placing D-branes on isolated singularities.

Our strategy will be to find an example of a geometry that has a non-isolated $A_1$ singularity
that at some point gets enhanced by an isolated singularity.
The fractional branes on the $A_1$ will provide the
$SU(N_f)$ symmetry;
they interact with $N$ branes at the
isolated singularity, that carry the $SU(N)$ color gauge group.
Such systems are easy to engineer.
The most basic examples are provided by
the generalized conifolds \cite{Gubser:1998ia}, the simplest of which is the
suspended pinch point singularity.\footnote{
The relevance of generalized conifolds and, in particular,
the suspended pinch point was stressed to us by Igor Klebanov.
See also \cite{KachruDSB}\cite{Amariti:2007am}\cite{Tatar:2007za}
for the earlier constructions of the
metastable SUSY breaking vacua in the generalized conifolds.
}

A similar mechanism of dynamical SUSY breaking for
the SPP singularity was previously considered in
\cite{Franco:2006es}:
SUSY is broken by the presence of D-branes on the deformed $A_1$
singularity.
The essential difference is that in our case
the $A_1$ singularity is deformed
without the conifold transitions within the SPP geometry.
In fact, we will show that the conifold transition is
responsible for SUSY restoration.

\subsection{D-branes at a deformed SPP singularity}

\begin{figure}[tp]
\begin{center}
\epsfig{figure=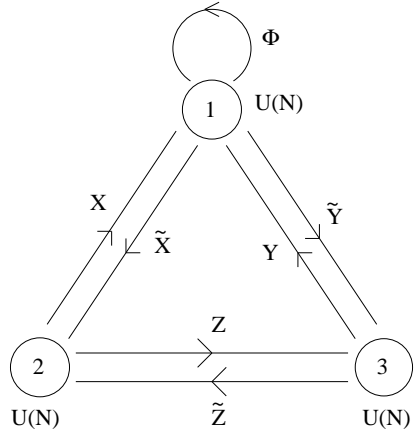,scale=0.6}
\vspace{-2mm}
\end{center}
\noindent
\caption{\it Quiver gauge theory for N D3-branes
at a suspended  pinch point singularity.}
\label{SPP1f}
\end{figure}

\noindent
The suspended pinch point (SPP) singularity may be obtained via a partial resolution of a $\Z_2\times \Z_2$ singularity  \cite{Morrison:1998cs}. It
is described by the following
complex equation in $\C^4$
\be
\qquad cd=a^2b, \qquad \qquad \quad \mbox{\small $(a,b,c,d) \in \C^4\, .$}
\ee
There is a $\C^2/\Z_2$ singularity along $b\neq 0$.
The quiver gauge theory for $N$ D3-branes at the SPP singularity
is shown in figure \ref{SPP1f}.
It was derived in \cite{Morrison:1998cs} by turning on an FI parameter $\xi$ in the
$\Z_2\times \Z_2$ quiver gauge theory, and working out the resulting symmetry
breaking pattern.
The superpotential of the SPP quiver gauge theory reads
\be\lb{SP1}
W=\Tr\left(\Phi(\td YY-\td XX)
+h (Z\td Z X\td X-\td ZZY\td Y)\right)
\ee
where $h$ is a dimensionful parameter
(related to the FI parameter via $h=\xi^{-1/2}$).

As a quick consistency check that this theory
corresponds to a stack of D3-branes on the SPP singularity,
consider the F-term equations for a {single} D3-brane.
The gauge invariant combinations of the fields are
\beaa\nonumber
a=\td XX=\td YY\quad & &  \quad c=X\td Y\td Z\\
\lb{deff}
b=Z\td Z\qquad\qquad \ & & \quad d=Y\td X Z
\eeaa
where we used the F-term equation for $\Phi$.
These quantities $(a,b,c,d)$ satisfy the constraint $cd=a^2b$,
which is the same as the equation for the SPP singularity.

Following our recipe as outlined in the introduction, we now deform the non-isolated $A_1$ singularity inside the SPP as follows
\be
\label{spd}
cd=a(a-\zeta)b.
\ee
This deformation removes the $A_1$ singularity,  replacing it by a
finite size 2-cycle. The deformed SPP geometry has two conifold singularities, located
at $a=0$ and $a=\zeta$,
with all other coordinates equal to zero.
In the field theory, the above deformation corresponds to adding an $F$-term
 of the form 
\bea\lb{Fterm2}
W_\zeta =- \zeta \Tr(\Phi 
- h \td ZZ).
\eea
This extra superpotential term is chosen such that the F-term equations for $\Phi$ and $Z$
\bea
\label{F2}
\td X X - \td Y Y -\zeta = 0, \quad & & \quad
 \td Z(\td Y Y - \td X X + \zeta) =0,
 \eea
are compatible.

The correspondence between (\ref{Fterm2}) and (\ref{spd}) is easily verified.
Again, consider the gauge theory on a single D3-brane. In view of the deformed F-term equation, the quantity $a$ now needs to be defined via
\bea
a=\td YY=\td XX+\zeta.
\eea
The constraint equation thus gets modified to 
$cd=a(a-\zeta)b$, which is the equation for the deformed SPP singularity.

As we increase $\zeta$, the two conifold singularities at $a=0$ and $a=\zeta$
become geometrically separated and the D-branes end up
on either of the two conifolds.
The field theory
should thus contain two copies of the conifold quiver gauge theory.
To verify this, consider the vacuum $Y=\td Y=\sqrt{\zeta}I$,
which solves both the F-term equations (\ref{F2}) and the D-term equations $|Y|^2-|\td Y|^2=0.$
These vevs break the gauge group $SU(N)_1\times SU(N)_3$ to
$SU(N)_{diag}$ and give a mass to the Higgs-Goldstone field
$Y_-=\frac{1}{\sqrt 2}(Y-\td Y).$
Substituting the remaining fields in the superpotential, one finds
that the fields $\Phi$ and $Y_+= \frac{1}{\sqrt 2}(Y+\td Y)$ are also massive.
The surviving massless  fields with the superpotential
\bea
W_{\rm con}=h(Z\td ZX\td X-\td ZZ\td XX)
\eea
reproduce the conifold quiver gauge theory.

In general, both $X$ and $Y$ have vevs and the D-branes split into
two stacks $N_1+N_2=N$ that live on the two conifolds.
Note, that the $Z$ field in (\ref{Fterm2}) corresponds to strings
stretching between the two conifolds.
The mass of this field is proportional to the length of the string
given by the size of the deformed two-cycle.


\subsection{Dynamical SUSY breaking}

\begin{figure}[t]
${}$ \ \ \ \ \ \ \ \ \ \ \ \  \parbox{15cm}{
\begin{flushleft}
\epsfig{figure=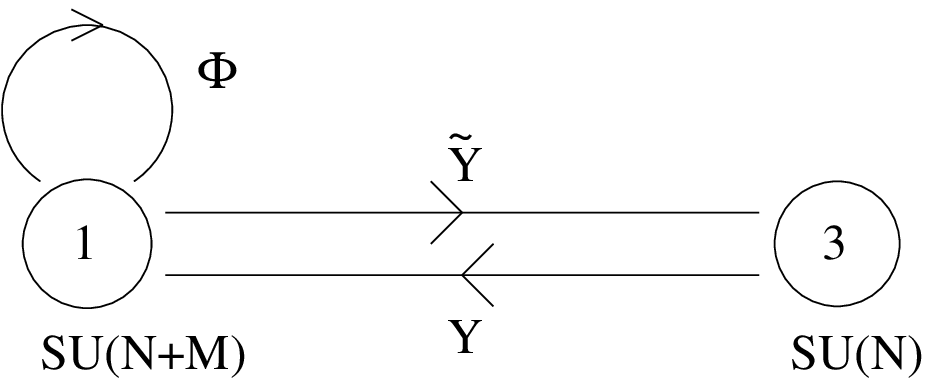,scale=0.6}
\end{flushleft}
\vspace{-4cm}
\hspace{1cm}
\begin{flushright}
\epsfig{figure=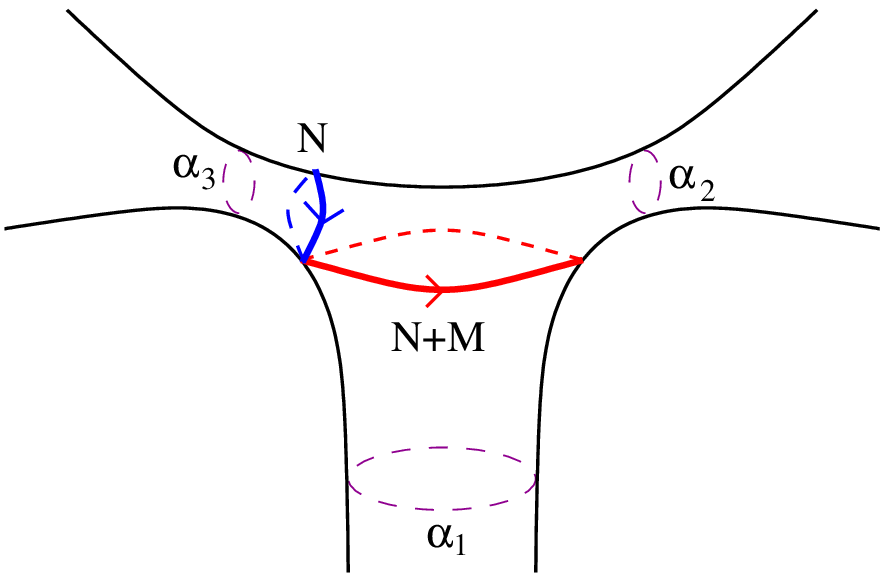,scale=0.75}
\vspace{4mm}
\end{flushright}}
\noindent
\caption{\it
A particular combination of fractional branes on the SPP singularity and
the corresponding quiver gauge theory that reproduce the ISS model.
The cycle $\al_1$ is a non-isolated two-cycle of the deformed $A_1$
singularity inside the SPP.
The cycles $\al_2$ and $\al_3$ denote the isolated two-cycles on the
two conifolds that remain after the deformation of the $A_1$
singularity.
The cycles satisfy $\al_1+\al_2+\al_3=0$.
The $N$ fractional branes wrapping $\al_3$ are supersymmetric.
The $N+M$ fractional branes wrapping $\al_1$ break SUSY.
This combination of fractional branes corresponds to zero vevs of
the bifundamental fields in the ISS.
}
\label{spp-geom1}
\end{figure}

\noindent
A straightforward way to generate dynamical SUSY breaking is to
reproduce the ISS model by placing some fractional branes on the SPP
singularity.
Suppose that there are $N_f=N+M$ fractional branes corresponding to node $1$ in figure \ref{SPP1f}, $N$ fractional branes
corresponding to node $3$, and no fractional branes at node
$2$.
The reduced quiver diagram is shown in figure \ref{spp-geom1}.
The superpotential for this quiver gauge theory is
\be\lb{isssu}
W=h \, \zeta\,
\Tr( \Phi) - h\, \Tr\bigl( \Phi Y\td Y\bigr)\, ,
\ee
which is the same as the ISS superpotential in the IR limit \cite{ISS}, with the $SU(N)$ identified
as the ``color'' group and $SU(N+M)$ as the ``flavor'' symmetry.
The only
difference between our gauge theory and the ISS system is that
the 
``flavor'' symmetry is gauged.
The corresponding gauge coupling is proportional to a certain period
of the B-field.
We can tune it to be small and treat the gauge group as
a global symmetry in the analysis of stability of the vacuum.%
\footnote{
In fact, the restriction on the coupling is not very strong, because
the SUSY breaking field $\Tr\Phi$ couples only to the bifundamental
fields $Y$, $\td Y$ through the superpotential (\ref{isssu})
(see also figure \ref{spp-geom1}).
Since the stabilization of the SUSY vacuum comes from the masses of
these bifundamental fields it is sufficient to require that the
corrections to the masses due to the gauge interactions are small
at the SUSY breaking scale.
}


An empty node in the quiver introduces some subtleties, since there
might be instabilities or flat directions at the last step of duality
cascade leading to this empty node.
In Appendix A we show that this quiver can be
obtained after one Seiberg duality from an SPP quiver without empty
nodes.

Recall that in the field theory SUSY is broken since
the F-term equations for $\Phi$
\be\lb{Fterm22}
Y \td Y=\zeta \, \mathds{1}_{N+M}
\ee
cannot be satisfied by the rank condition.
In the vacuum where
\be
\lb{newyy}
Y \td Y  = \zeta \, \mathds{1}_{N},
\ee
the $SU(N)_1\times SU(N+M)_3$ gauge symmetry is broken to
$SU(N)_{diag} \times SU(M)_3$. The
 superpotential for the remaining $M\times M$ part of the adjoint field reduces to
 the Polonyi form
\bea
W = h\, \zeta \,
\Tr_M( \Phi).
\eea
The metastable ground state thus has a vacuum energy proportional to $M h^2 \zeta^2$.

We can interpret the SUSY breaking vacuum on the geometric side as follows.
Our system contains  $N$ fractional
branes that wrap one of the conifolds inside the deformed SPP singularity,
and $(N+M)$ fractional branes that wrap the 2-cycle of the deformed $A_1$.
The $\Phi=0$ vacuum corresponds to putting all
the $(N+M)$ fractional branes on top of the $N$ branes at the
conifold (see fig. \ref{spp-geom1}).

\begin{figure}[t]
\begin{center}
\epsfig{figure=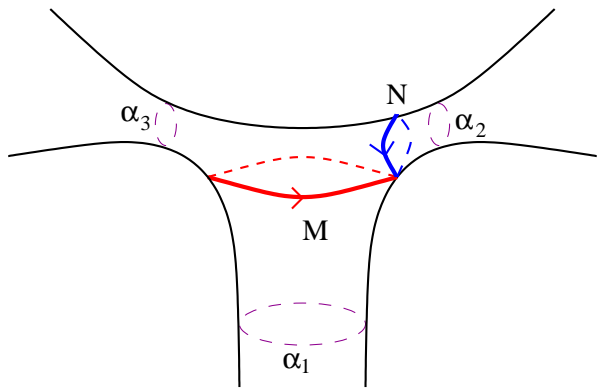,scale=0.88}
\vspace{-3mm}
\end{center}
\noindent
\caption{\it In the metastable vacuum. $N$ supersymmetric fractional branes
wrap the $-\al_2$ cycle of the second conifold.
The remaining $M$ fractional branes
wrap the non-isolated cycle $\al_1=-\al_2-\al_3$ and are weakly
bound to the $N$ branes at the conifold.
This configuration of fractional branes is obtained from the
configuration in figure \ref{spp-geom1}
by giving vevs to the bifundamental fields.}
\label{meta}
\end{figure}

The $Y$ modes represent the massless ground states of
the open strings that connect the two types of branes.
The non-zero expectation value (\ref{newyy}) for $Y\td Y$ corresponds to
a condensate of these massless strings between $N$
branes wrapping $\al_3$ and $N$ branes wrapping $\al_1=-\al_2-\al_3$.
As a result of condensation, these two stacks of $N$
fractional branes recombine into $N$ fractional branes wrapping $-\al_2$
at the second conifold.
The remaining $M$ fractional
branes around the deformed $A_1$  end up in a non-supersymmetric state. The diagonal entries
of the $M\! \times \! M$  block in $\Phi$ parameterize the motion of
the $M$ branes along the deformed non-isolated singularity.
The corresponding configuration of branes is represented in figure
\ref{meta}.

 The stability of the SUSY breaking vacuum is a quantum effect in the
field theory--- there are pseudo-moduli that acquire a stabilizing potential
at one loop \cite{ISS}. In the D-brane picture this should correspond
to the back reaction of the branes that makes the two-cycle at the deformed $A_1$
singularity grow as one moves away from the conifold.
(Alternatively one can think about a weak attraction between the
branes.)
It would be interesting to derive this directly from
SUGRA equations, since it would complete the geometric
evidence for the existence of the SUSY breaking vacuum.

\subsection{SUSY restoration}

\noindent
Let us discuss SUSY restoration in this setup.
The SUSY vacuum is found by  separating the $(N+M)$ fractional branes on the deformed $\C^2/\Z_2$ singularity from
the $N$ fractional branes at the conifold. This separation amounts to giving a vev to $\Phi$.
Initially this costs energy. The fields $Y$ and $\td Y$ become massive.
Below their mass scale, the theory on the $N$ fractional branes at the conifold becomes strongly
coupled and develops a gaugino condensate. This condensate deforms the
conifold singularity,  and generates an extra term in the
superpotential for $\Phi$ that eventually restores SUSY.

On the gauge theory side, the SUSY restoring
superpotential term arises due to the fact that
the value of the gaugino condensate depends
on the masses of $Y$ and $\td Y$, and these in turn depend on the vev of
$\Phi$.
As a result \cite{ISS},
the gaugino condensation modifies the superpotential for $\Phi$ to (here  $N_f = N+M$)
\bea\lb{effsuper3}
W_{low}=N\Bigl(h^{N_f}\Ld_m^{-(N_f-3N)}\det\Phi\Bigr)^{1/N} - h\zeta \, \Tr\Phi.
\eea
Due to the extra term, the F-term equations
\be
\frac{\p W_{low}}{\p\Phi}=0
\ee
can be solved. In fact there are $N_f\! -N\! =\! M$ SUSY vacua $\Phi\! =\! \Phi_k$, with $k=1,..\, , M$.   

On the geometric side, the SUSY vacuum is interpreted as the ground
state of $N+M$ fractional branes in the presence of a deformed
conifold singularity.
Suppose that the deformed conifold is the one located at $a=\zeta$.
One can describe the situation after the geometric
transition by the following equation
\bea\lb{geometry3}
cd=a((a-\zeta)b+\e)\, .
\eea
The original conifold singularity at $a=\zeta$ is now a smooth point in the
geometry. However, a new singularity has appeared in the form of an
undeformed conifold at $a=c=d=0$ and $b=\e/\zeta$.
The D5-branes that were originally stretching between $a=0$ and
$a=\zeta$ can thus collapse to a supersymmetric state by wrapping
the zero-size 2-cycle of the undeformed conifold. This process is the geometric
manifestation of SUSY restoration
in the underlying ISS gauge theory.%
\footnote{A similar mechanism of SUSY restoration in the case of SPP
singularity was anticipated in \cite{Franco:2005zu}
}

\begin{figure}[t]
\begin{center}
\epsfig{figure=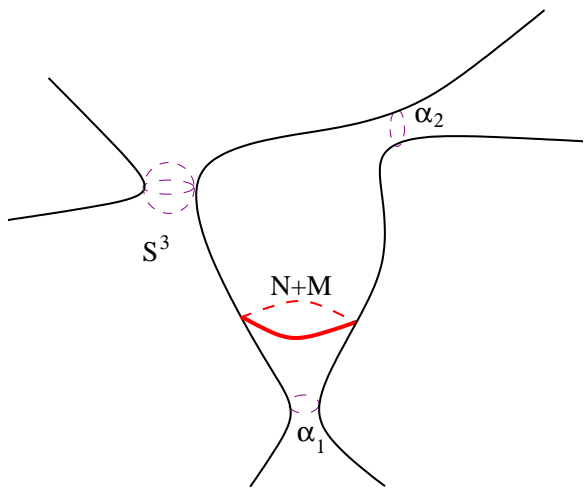,scale=0.85}
\vspace{-3mm}
\end{center}
\noindent
\caption{\it To reach the supersymmetric ground state, the $N+M$ fractional branes on the $A_1$ 2-cycle
move away from the conifold. The $N$ fractional branes on the conifold then drive the geometric
transition: the two-cycle $\al_3$ is replaced by the three sphere
$S^3$.
After the transition, the size of the $A_1$ 2-cycle reaches a zero minimum at a new
conifold singularity (indicated by the position of $\al_1$).
}
\label{restore}
\end{figure}

Using the geometric dual description, it is possible to rederive the field
theory superpotential (\ref{effsuper3}) and even compute higher-order corrections.
The calculation goes as follows,  \cite{Aganagic:2007py}.
Let us rewrite the geometry (\ref{geometry3})
as: 
\be
\label{zxx}
uv=(z-x)((z+x)(z-x-\zeta)+\e)\, ,
\ee
where $z-x=a$, $z+x=b$.
Also it is useful to introduce the following notation
\beaa
\label{zzz}
\nonumber
z_1(x)=x \qquad & & \\[-4mm]
\lb{defz} & & \qquad
\td z_2(x)=\textstyle {\zeta}/{2}-\sqrt{(x+\zeta/2)^2-\e}
\nonumber\\[-2mm]
z_2(x)=-x \qquad & & \\[-2mm]
& & \qquad \td z_3(x)=\textstyle {\zeta}/{2}+\sqrt{(x+\zeta/2)^2-\e} \nonumber \\[-4mm]
z_3(x)=x+\zeta \qquad & & \nonumber
\eeaa
The conifold singularity is at $z=x={\e}/{2\zeta} \equiv x_*$.
If initially the fractional D-branes on the deformed $A_1$ were stretching between
$z_1(x)$ and $z_3(x)$, then after the geometric transition, they stretch between
$z_1(x)$ and $\td z_3(x)$.  They can minimize their energy by moving (or tunneling)
to the conifold
singularity at $z=z_1(x_*)=\td z_3(x_*)$. 

\begin{figure}[t]
\begin{center}
\epsfig{figure=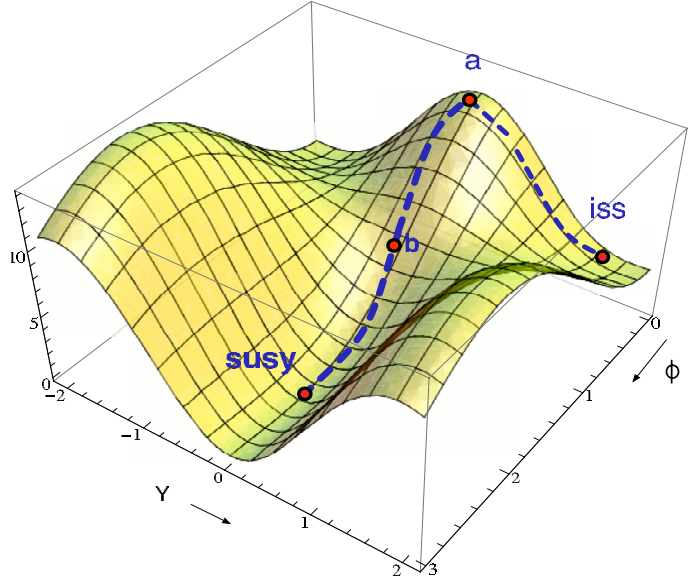,scale=1.3}
\vspace{-6mm}
\end{center}
\noindent
\caption{\it The gauge theory potential as a function of the vevs of the
adjoint $\Phi$ and bi-fundamental $Y$.
Suppose we start at point $\bf a$ corresponding to the situation in
fig. \ref{spp-geom1} with the zero vevs of $\Phi$ and $Y\td Y$.
This point is unstable and there are two possibilities.
If the bifundamental fields $Y\td Y$ get a vev, then we end up in the
metastable ISS vacuum in fig. \ref{meta}.
If the adjoint field $\Phi$ gets a vev, then we follow the
path to the SUSY vacuum (an intermediate point on the SUSY restoring
path is shown in fig. \ref{restore}).}
\label{isspot}
\end{figure}

For the geometric derivation of the superpotential, we take
the deformation parameter $\e$ to be dynamical, and related to the
gaugino condensate via $\e=2S$.%
\footnote{The constant $2$ appears due to the
consistency conditions between the geometric derivation of the
superpotential and the KS superpotential for the conifold.
}
We also identify  $\Phi$ with the location $x$
of the D5-branes relative to the (deformed) conifold at $a=\zeta$.
The superpotential for the gaugino condensate together with the
adjoint field is \cite{Aganagic:2007py}
\bea\lb{Spoten}
W(S,\Phi)=NS(\log\frac{S}{\Ld^3}-1)+\frac{t}{g_s}S+\td W(\Phi,S)\, .
\eea
The first two terms comprise the familiar GVW superpotential
\cite{GVW}
$W = \int \Omega \wedge G_{\it 3}$
evaluated for the deformed conifold supported by $N$ units of RR 3-form flux
\cite{geomtrans}\cite{Giddings:2001yu}.
The last term  $\td W(\Phi,S)$ has a closely related, and
equally beautiful, geometric characterization in terms
of the integral of holomorphic 3-form
\bea
\label{nice}
\td W(\Phi,S) = \int_\Gamma \Omega
\eea
over a three chain $\Gamma$ bounded by the 2-cycle wrapped by  the D5 brane.%
\footnote{This contribution
to the superpotential is easily understood from the perspective of the GVW  superpotential.
The D5-brane is an electric source for the RR 6-form potential $C_{\it 6}$, and a magnetic source
for the RR 3-form field strength $F_{\it 3} = dC_{\it 2}$. If the D5 would traverse some 3-cycle $\mathsf A$,
this process will induce a jump by one unit in the $F_3$-flux through the 3-cycle $\mathsf B$ dual to $\mathsf A$, and thereby a corresponding jump in the GVW superpotential.
Continuity of the overall superpotential during this process dictates that the D5-brane
contribution must take the form (\ref{nice}).} Following \cite{Aganagic:2007py}, we can
reduce the integral (\ref{nice}) for our geometry (\ref{zxx}) to an indefinite 1-d integral
\be
\label{geom}
\td W(x)=\int (\td z_3(x)-z_1(x))dx
\ee
with $z_1(x)$ and $\td z_3(x)$ given in (\ref{zzz}).

Let us show that the geometric expression (\ref{geom}) reproduces the gauge theory
superpotential.
In the appropriate limit,  $x>\!\!>\e ,\zeta$, we find from (\ref{geom})
\be
\td W(S,\Phi)=\zeta\Tr\Phi - S\log\bigl({\Phi}/{\Ld_m}\bigr)\, .
\ee
Here we identify $(x,\epsilon)$ with $(\Phi,2S)$, and use the
integration constant to introduce a scale $\Ld_m$. Physically, $\Lambda_m$ sets the scale of
 the Landau pole for the IR free theory with $3N<N_f$.
Minimizing (\ref{Spoten}) with respect to $S$ we find
\be
S=\left(\Ld_m^3\det\bigl({\Phi}/{\Ld_m}\bigr)\right)^\frac{1}{N}
\ee
If we substitute $S$ back in (\ref{Spoten}), we get
exactly (\ref{effsuper3}) (up to an overall sign and after the redefinition
$\Phi\ra h\Phi$).
By expanding the full geometric
expression (\ref{geom}) to higher orders, one can similarly extract the multi-instanton corrections
to the superpotential.

Our system in fact has other supersymmetric vacua besides the one just exhibited.
These arise because, unlike the ISS-system, the flavor symmetry is gauged.
If we move $M$ fractional branes away from the conifold
singularity in figure \ref{meta}, then the $N$ fractional branes wrapping
the conifold 2-cycle $\al_2$  may also induce a geometric transition.
As in the above discussion, this transition also restores SUSY. For a suitable
choice of couplings, the extra SUSY vacuum lies farther away than the one
considered above.
The ISS regime arises when the coupling of the initial \lq\lq color" $SU(N)$
gauge group is sufficiently bigger than the coupling of the gauged
\lq\lq flavor" group $SU(N+M)$, $g_3\gg g_1$.
(Note that after the symmetry breaking, the coupling of
$SU(N)_{diag}\subset SU(N)\times SU(N+M)$ is
of order $g_1$.)
In this section we assumed that we are in this ISS regime.


\section{Type IIA dual of the SPP singularity}

\noindent
In this section we present the type IIA dual of our discussion of D-branes at the SPP
singularity. In particular, we study the
F-term deformations in the corresponding system of NS-branes and D-branes and
prove that the IIA dual of the SPP singularity is equivalent to the
known IIA representations of ISS \cite{Ooguri:2006bg, IIAfollow, Giveon:2007fk}.

\begin{figure}[hbtp]
\begin{center}
\epsfig{figure=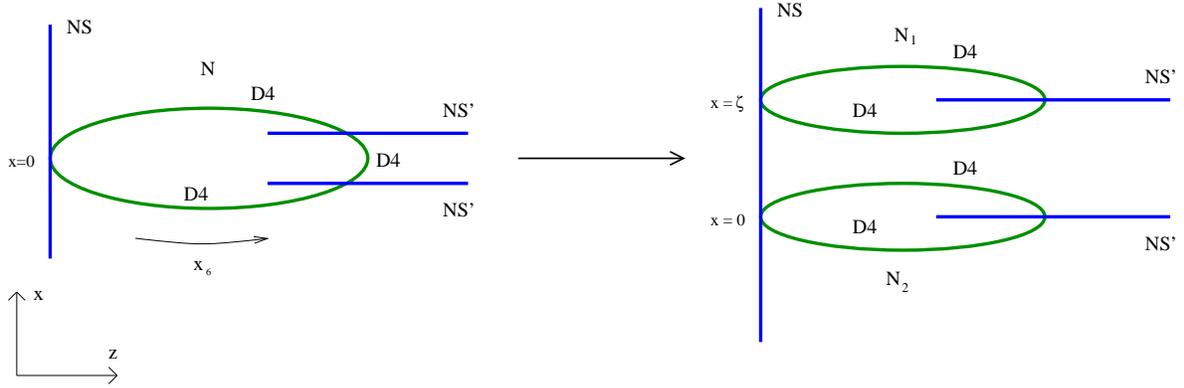,scale=0.7}
\end{center}
\noindent \caption{\it On the left there are N D4-branes on the \lq\lq SPP
singularity". After the addition of the F-term $\zeta$, the SPP
singularity is transformed to two conifolds at $x=0$ and $x=\zeta$.
The $N$ D4-branes split into $N_1+N_2=N$ D4-branes at the two
conifolds. (The D4-branes are green and the NS branes are blue.)
}
\label{NS-SPP1}
\end{figure}

D-branes at singularities of CY manifolds in IIB are T-dual to
D-branes stretching between NS-branes in type IIA
\cite{Ooguri:1995wj}\cite{Uranga:1998vf}. Consider $N$ D3-branes at
the SPP singularity described by the following equation in $\C^4$
\be\lb{SPP4}
uv=x^2z.
\ee
The resulting space has 6 real dimensions $(x_4,\ldots,x_9)$.
Denote $x=x_4+ix_5$ and $z=x_8+ix_9$. For $v\neq 0$ one can solve equation
(\ref{SPP4}) for $u$. Let $v=re^{i\vp}$ and denote $x_6=\vp$,
$x_7=r$. After T-duality in the compact dimension $x_6$, we get
the configuration of NS branes (blue) and D4-branes (green) in type
IIA (this configuration is depicted in figure \ref{NS-SPP1} on the
left).

The zeros of polynomials on the right hand side of (\ref{SPP4})
 represent the intersection of
NS-branes with the circle in $x_6$. There is one NS brane at $z=0$
and two NS$'$ branes at $x=0$ (we use the prime to distinguish
the two NS branes at $x=0$ from the NS brane at $z=0$). The NS branes span the
following dimensions
\bea
NS\;\;\;(0\;1\;2\;3\;4\;5)\\
NS'\;\;\;(0\;1\;2\;3\;8\;9)
\eea
The D4-brane between the two NS$'$ branes can freely move in the $z$
direction. This corresponds to the motion of the fractional
D3-branes along the line of $Z_2$ singularities in the $z$ direction
of (\ref{SPP4}).
The length of the D4 brane in $x^6$ is mapped, via T-duality, to the period of the B-field on the corresponding shrunken $\mathbb{P}^1$:
\begin{equation}\label{periodmap}
\Delta x^6=\int_{\mathbb{P}^1}B\sim \frac{4\pi}{g^2}
\end{equation}

The corresponding field theory is the same as the type IIB quiver
gauge theory (\ref{SP1}). As we have shown earlier the F-term
deformation (\ref{Fterm2}) corresponds to the deformation of the
$Z_2$ singularity in the SPP
\be
uv=x(x-\zeta)z
\ee
In the IIA dual picture this corresponds to moving one of the NS'
branes from $x=0$ to $x=\zeta$. This theory has two conifold points:
at $x=0$ and at $x=\zeta$. The corresponding configuration of branes
is shown in figure \ref{NS-SPP1} on the right.

\begin{figure}[hbtp]
\begin{center}
\epsfig{figure=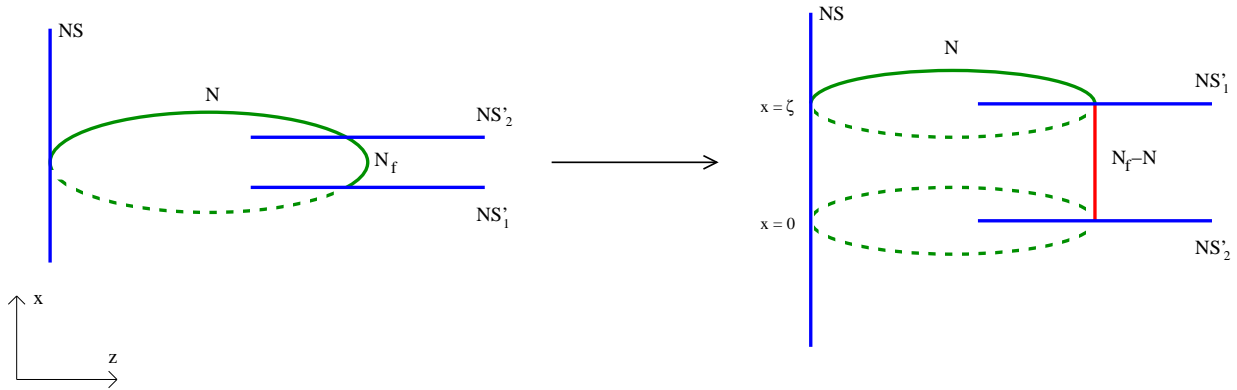,scale=0.7}
\end{center}
\noindent \caption{\it On the left there are $N_f$ D4-branes
stretching between $NS_1'$ and $NS_2'$ and $N$ D4-branes stretching
between $NS_2'$ and $NS$. After turning on the F-term, $\zeta$, there
are $N$ supersymmetric D4-branes (green) stretching between $NS_1'$
and $NS$ and $(N_f-N)$ D4-branes (red) that have non zero size in
the $x$ direction and violate the SUSY. The dashed lines represent
the empty cycles in the geometry (empty nodes in the corresponding
quivers). } \label{NS-SPP2}
\end{figure}

To get the ISS vacuum we take $N_f$ D4-branes between the two NS$'$
branes and $N$ D4-branes between $NS_2'$ and $NS$ such that $N_f>N$.
The corresponding superpotential is
\be
W=\Tr(\zeta\Phi-\Phi\vp\td\vp)
\ee
The F-term equations for the $\Phi$ fields are
\be
\vp\td\vp=\zeta I_{N_f\times N_f}.
\ee
The fields $\vp$ and $\td\vp$ acquire vevs and break the gauge group
as $SU(N_f)\times SU(N)\lra SU(N)_{diag}\times SU(N_f-N)$. This
corresponds to recombination of the D4-branes shown in figure
\ref{NS-SPP2}.
The SUSY breaking is due to the $(N_f-N)$ D4-branes
stretching a finite distance between $x=0$ and $x=\zeta$: the
tension of these branes creates the vacuum energy.
We note, that this configuration of NS-branes and D4-branes is
closely related to the constructions of
\cite{Giveon:2007fk}
where the $SU(N_f)$ symmetry is slightly gauged
compared to the earlier constructions
\cite{Ooguri:2006bg, IIAfollow}
where the $SU(N_f)$ is a flavor symmetry.

\section{F-term via a Deformed Moduli Space}

\noindent
In the previous sections we introduced the F-terms by hand,
assuming that they are generated
somewhere else in the geometry and are not affected by the local field
theory (see, e.g., the constructions in \cite{Aganagic:2007py}).
In this section we consider an example of F-term generation in the
local field theory by a quantum modified moduli space analogous
to the Intriligator-Thomas-Izawa-Yanagida model \cite{IT,IY}. Our setup is related to the M-theory example considered in \cite{M5DSB}.

In order to obtain an ITIY-like model we consider the
deformed $A_3$ singularity in IIB string theory:
\be\label{A3eqn}
uv=x^2z^2.
\ee
Recall that the $C^2/Z_4=A_3$ singularity has the equation $uv=x^4$
in $\C^3$.
The corresponding quiver gauge theory \cite{Douglas:1996sw}
for $N$ D3-branes at the $A_3$ singularity has four $U(N)$ gauge
groups, four $\N=2$ hypermultiplets in bifundamental
representations of the gauge groups, and four adjoint fields.

The deformation (\ref{A3eqn}) corresponds to giving the masses to
two adjoint fields on opposite nodes of the $C^2/Z_4$ quiver.
A general derivation of the correspondence between the
geometric deformations and the superpotential for the
adjoint fields can be found in \cite{Cachazo:2001sg}.
Intuitively, an adjoint field gets a mass if the corresponding
fractional brane wraps a collapsed two-cycle that has a non-zero
volume away from $x=z=0$.
After integrating out the massive adjoint fields,
the remaining fields are the
four $U(N)$ gauge groups with bifundamental
matter between them and two adjoint fields corresponding to the
non-isolated $Z_2$ singularities at $u=v=x=0$ and $u=v=z=0$.

Next, let us add an O3 plane located at $u=v=x=z=0$.
We take the action of the O3 plane to be the
same as in \cite{O3KS}:
\begin{equation}\label{O3action}
u\to v, \ \ v\to u, \ \ x\to -x, \ \ z\to -z
\end{equation}
The $U(N)$ gauge groups become $SO(2N+2)$ and $Sp(N)$.

\begin{figure}[hbtp]
\begin{center}
\epsfig{figure=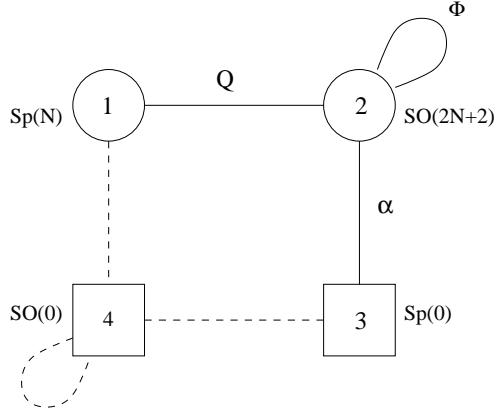,scale=0.75} \vspace{-3mm}
\end{center}
\noindent
\caption{\it
Quiver gauge theory that reproduces the ITIY model.
It is obtained by putting some fractional branes on an orientifold
of the deformed $A_3$ singularity.
The circles represent the occupied nodes, while the squares
correspond to the empty nodes in the quiver.
The field $Q$ is in the bifundamental representation of $Sp(N)\times
SO(2N+2)$.
$\al$ denotes fermionic zero modes of
the D-instantons wrapping the $Sp(0)$ node.
}
\label{A3}
\end{figure}

To generate the ITIY model, we occupy two out of the four nodes
in the quiver.
The corresponding quiver gauge theory is shown in figure
\ref{A3}.
The $N$ fractional branes corresponding to node 1 give rise to an $Sp(N)$ gauge
theory with dynamical scale $\Lambda$, while the $N+1$ fractional
branes corresponding to node 2 realize an $SO(2N+2)$ theory with dynamical
scale $\Lambda'$.
In our example, the beta function for the $Sp(N)$ gauge group is bigger than for
$SO(2N+2)$, i.e. the $Sp(N)$ gauge group confines first.
We assume that $\Lambda\gg\Lambda'$
and treat the weakly coupled $SO(2N+2)$ symmetry as global.

The tree-level superpotential is inherited from the $\mathbb{C}^2/Z_4$ cubic
superpotential
\begin{equation}\label{W}
W=h\Phi_{ij}Q^iQ^j
\end{equation}
where $\Phi$ is an adjoint of $SO(2N+2)$ and
the quarks, $Q$, transform as bifundamentals of
$Sp(N)\times SO(2N+2)$.

Denote the mesons of the $Sp(N)$ gauge group by $M^{ij}=Q^iQ^j$.
After the confinement of $Sp(N)$, the theory has a quantum-deformed
moduli space of vacua
\be\lb{Quantummodspace}
{\rm Pf}M=\Ld^{2N+2}
\ee
The superpotential (\ref{W}) then becomes
\be\lb{ITIYsup}
\td W=h\Phi M + \ld ({\rm Pf} M -\Ld^{2N+2}).
\ee
where $\ld$ is the Lagrange multiplier imposing the constraint
(\ref{Quantummodspace}).

SUSY is broken since the F-term equations for the $\Phi$ field
cannot be satisfied.
Indeed, the deformed moduli space guarantees that
\begin{equation}\label{FPhi}
-F_{\Phi}^{\dagger}=
M\sim\Lambda^2\ne 0.
\end{equation}
Note that we needed to introduce the O3 plane in order to
(dynamically) break SUSY since otherwise
we would have to take baryonic directions $B, \tilde{B}$ into account in
(\ref{Quantummodspace}). In the
absence of competing effects, the baryons are tachyonic and so our
potential would take us to zero vev for $M$, thus allowing the
system to relax to a SUSY groundstate.

In order to get a geometric
interpretation of the SUSY breaking,
let us solve the F-term equations for the $\ld$ and $M$ fields
\beaa\label{FlambdaFterm}
{\rm Pf}M-\Lambda^{2N+2}=0;\\
\nonumber
h\Phi_{ij}+{\lambda}{\rm Pf}M\cdot M_{ij}^{-1}=0.
\eeaa
Then, the superpotential for $\Phi$ reads
\be\label{effW}
\td W=2h\Ld^2 (N+1) ({\rm Pf}\Phi)^{\frac{1}{N+1}}.
\ee
Any $\Phi$ can be obtained by an $SO(2N+2)$ rotation from a given
element $\Phi_0$, $\Phi=O \Phi_0 O^T$, where we take
\be\label{phi0}
\Phi_0=\left(
\ba{cc}
0 & R\\
-R & 0
\ea
\right)
\ee
with
\be
R={\rm diag}(r_1,...,r_{N+1})
\ee
The anti-symmetric form of $\Phi$ is due to the orientifold projection. Now, plugging (\ref{phi0}) into (\ref{effW}) and extremizing the resulting potential, we see that
\be\lb{effpot}
V=4h^2\Lambda^4\left(\sum_i\frac{1}{|r_i|^2}\right)\prod_j|r_j|^{\frac{2}{N+1}}\ge4h^2\Lambda^4(N+1)
\ee
with the inequality saturated for $r_1=...=r_{N+1}$, i.e.
\be\lb{saturate}
\Phi_0=r\left(
\ba{cc}
0 & \mathds{1}_{N+1}\\
-\mathds{1}_{N+1} & 0
\ea
\right).
\ee
Then ${\rm Pf}\Phi=r^{N+1}$ and
\be
\td W=2h\Ld^2(N+1)r.
\ee
In other words, this is a Polonyi model in the flat $r$ direction with a set of Goldstone bosons parameterizing the space of broken symmetries $SO(2N+2)/U(N+1)$. In fact, these goldstone bosons will get eaten at the scale $\Lambda$ since
\be\lb{Mfix}
M=\Lambda^2\left(
\ba{cc}
0 & \mathds{1}_{N+1}\\
-\mathds{1}_{N+1} & 0
\ea
\right).
\ee
as a consequence of satisfying the F-term equations in (\ref{FlambdaFterm}) (this holds for $\forall r\ne 0$ and therefore holds in the limit $r\to 0$).

Now, by construction, $r$ is uncharged under the $U(N+1)$ group of remaining symmetries. Hence, in particular, we can treat $r$ as a center of mass coordinate of the D-brane system.  Then, in analogy with the previous sections, we interpret this superpotential
as coming from the complex deformation of the singularity
\begin{equation}\label{deformed}
uv=(z-{h\Lambda^2})(z+h\Lambda^2)x^2
\end{equation}
Here we take the deformation to be invariant under the O-plane
action.
In the case of the ITIY model, this geometric interpretation has an
important limitation.
In the previous constructions we assumed that the deformation
parameter $\zeta$ is a vev of some field that has a mass much bigger
than the scale of $\zeta$, i.e. that we can decouple its dynamics
from the D-brane dynamics.
For the ITIY, the mass of the $M$ fields is proportional to $\Ld$,
i.e. the dynamics of $M$ start to play a role already at the SUSY
breaking scale.
In other words, the geometric formula (\ref{deformed}) should be
trusted only for $x,z,u,v\ll h\Ld^2$.

Let us now show that SUSY is restored in this model by
contributions from the D-instantons wrapping the empty nodes in
quiver \ref{A3}. The presence of empty nodes
seems rather generic in constructions of the ITIY model from D-branes at singularities.
The presence of the O3-plane then allows non trivial D-instanton contributions to the superpotential.%
\footnote{Additional non-perturbative effects in the $U(N+1)\subset SO(2N+2)$ theory are small provided we take $\Lambda'$ sufficiently small.}
In the case of the $Sp(0)$ node, the \lq +' orientifold
projection lifts the additional zero modes of the D-instanton and
allows it to contribute to the superpotential \cite{Lifting,
Dinstanton, MoreDinstanton} (the corresponding D-instanton zero modes are represented by $\al$
in figure \ref{A3}),
while the \lq -' sign of the projection on the $SO(0)$
node does not lift the extra zero modes and so no contribution to
the superpotential is expected from that node.

Integrating out the fermionic zero modes, $\alpha_i$, resulting from
a Euclidean D1 brane wrapping node 3 gives rise to an exponentially
suppressed deformation of the superpotential:
\begin{equation}\label{FullW}
W=h\Phi M + \epsilon {\rm Pf}\Phi
\end{equation}
where the suppression factor is given by:
\begin{equation}\label{epsilon}
\epsilon\sim e^{-t/g_s}
\end{equation}
with $t$ the period of $B^{\rm NS}+ig_sB^{\rm RR}$ on the
corresponding shrunken 2-cycle. Note that since $\Phi$ has $Sp(N)$
non-anomalous R-charge $+2$, the second term in (\ref{FullW}) breaks
the R-symmetry and SUSY will be restored for $\Phi$ that satisfies:
\begin{equation}\label{SUSYrestore}
\Phi\sim\frac{\Lambda^{2+2/N}}{h\epsilon^{1/N}}M^{-1}\sim
\left(\frac{\Lambda^{2}}{h^N\epsilon}\right)^{\frac{1}{N}}
\end{equation}
Since $\epsilon$ is parametrically small, we can take it such that
the model is rendered metastable. For $\Phi$ near the origin,
however, the stringy instanton contribution will be dominated by the
F-term induced by the $Sp(N)$ quantum deformed moduli space. Based
on the form of the D-instanton contribution, we are tempted to
identify this term with a geometrical transition in (\ref{deformed}). Formally, we can do
this and maintain compatibility with the orientifold projection in the limit in which $\Lambda\to 0$
(this reflects the fact that SUSY restoration occurs in an entirely different regime of field space
$\Phi\gg\Ld$):
\begin{equation}\label{geotrans}
uv=xz(xz-s)
\end{equation}

The stability of the SUSY breaking vacuum can be analyzed similarly
to \cite{CalculableDSB}.
The field $r$ introduced in (\ref{saturate}) is a pseudo-modulus.
This pseudo-modulus is lifted upward by corrections to the potential
leaving a metastable SUSY-breaking vacuum at the origin,
$
\Phi=0.
$

One might be worried that contributions from the gauge fields could destabilize the vacuum. The first thing to note is that the $r$ field is not charged under the subgroup
$U(N+1)\subset SO(2N+2)$ unbroken below $\Ld$. The contributions to the potential
from the broken $SO(2N+2)/U(N+1)$ gauge sector can be neglected if the
corresponding coupling is smaller than the coupling of the matter fields,
$g\ll h$, which can be arranged via an appropriate geometric tuning.%
\footnote{
The couplings of the gauge fields can be tuned by changing the
periods of the $B$-field.
If we had an accidental $\N=2$ supersymmetry, then the couplings $g$
and $h$ would be related, but in our $\N=1$ setup they are not protected
against independent changes.
}


\section{Concluding remarks}

\noindent
In this paper we presented a simple geometric criterion for the
existence of a meta-stable F-term SUSY breaking vacuum in
world-volume gauge theories on D-branes. We showed that  the basic
ingredients of the ISS theory can be realized by placing fractional
D-branes on a slightly deformed non-isolated singularity passing
through an isolated singularity. We characterized both the
meta-stable non-SUSY and stable SUSY vacuum states.

A gap in our study, and an important direction to be explored, is
the detailed supergravity analysis of the SUSY breaking vacuum. On
the field theory side, the one-loop corrections to the
potential are crucial for lifting the classical degeneracy and
stabilizing the meta-stable vacuum. In the D-brane picture this
corresponds to a weak attraction between the $N$ D-branes at the
isolated singularity and the $M$ D-branes at the non-isolated $A_1$
singularity. This attraction presumably  arises due to some back
reaction that slightly deforms the 2-cycle of the $A_1$
singularity, such that its area is minimized near the isolated
singularity.

Our construction may be used to introduce SUSY breaking in
phenomenological models involving D-branes at singularities of CY
manifolds. For example, take the construction of an SM-like theory
in terms of D-branes on a del Pezzo 8 singularity considered in
\cite{delPezzo8}. As argued in \cite{delPezzo8}, the symmetry
breaking towards the SM requires the formation of an $A_2$
singularity on the del Pezzo 8 surface. This $A_2$ lifts to a
non-isolated singularity on the cone over del Pezzo. The results in
this paper suggest that, if we slightly deform this non-isolated
singularity and put a suitable collection of fractional branes on
it, we can engineer a SUSY breaking hidden sector, with charged
matter that interacts with the SM part of the quiver. In this way we
may be able to build a semi-realistic phenomenological model.

\bigskip
\bigskip

\noindent
{\large \bf Acknowledgments.}
\medskip

\noindent
The authors are thankful to Sebastian Franco, Shamit Kachru, Igor Klebanov,
David Kutasov, Christopher Herzog,
Diego Rodr\'iguez-G\'omez, Nathan Seiberg, Eva Silverstein,
and Martijn Wijnholt for valuable discussions.
This work is supported by the Russian Foundation of Basic Research
under grant RFBR 06-02-17383 (DM), an NSF Graduate Research
Fellowship (MB), and by the National Science Foundation under grant
PHY-0243680. Any opinions, findings, and conclusions or
recommendations expressed in  this material are those of the authors
and do not necessarily reflect the views of the National Science
Foundation.

\newpage

\append{ISS quiver via an RG cascade}

\begin{figure}[t]
\begin{center}
\epsfig{figure=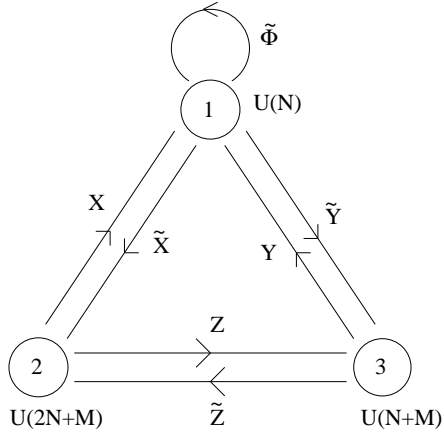,scale=0.6}
\vspace{-3mm}
\end{center}
\noindent
\caption{\it Quiver gauge theory of the fractional brane configuration on the SPP singularity
that reduces to ISS model after confinement of the $SU(2N+M)$ gauge
group at node 2.}
\label{SPP2}
\end{figure}

\noindent
In this appendix we show that the ISS quiver in figure \ref{spp-geom1}
can be obtained after one Seiberg duality from an SPP quiver
in figure \ref{SPP2}.

This quiver is obtained from the quiver in figure
\ref{SPP1f} by adding $M$ fractional branes to node 3, and
$N+M$ fractional branes to node 2, so that the
respective ranks of the gauge groups become $N+M$ and $2N+M$.
Note, that this theory has
an infinite duality cascade that increases the ranks of the gauge groups,
i.e. we can suppose that we start in the UV with some big ranks of
the gauge groups and after a number of duality steps arrive at
quiver \ref{SPP2}.
Let us show that after one more duality at node 2 we reproduce the
ISS model.

The theory has gauge group $U(N)\times U(2N+M)\times U(N+M)$,
one adjoint under $U(N)$ and three vector-like pairs of bi-fundamentals.
The superpotential is given by the sum of (\ref{SP1}) and (\ref{Fterm2})
\be
W=\Tr\left(-\zeta\td\Phi+\td\Phi(\td YY-\td XX)
+h(Z\td Z X\td X-\td ZZ Y\td Y+\zeta \td ZZ)\right).
\ee

The $SU(2N+M)$ gauge group confines first.
This gauge group has $N_f=N_c$ and thus the gauge group after the
Seiberg duality is $U(N)\times U(N+M)$.
The two $U(1)$ factors can be represented as the
overall $U(1)$ (that decouples) and the non-anomalous $U(1)_B$ gauge
group. Denote the meson fields as
$M_{xx}=\td XX$,
$M_{xz}=\td XZ$,
$M_{zx}=\td ZX$, and
$M_{zz}=\td ZZ$.
In addition there are two baryons $A$ and $B$.
After the Seiberg duality, the superpotential is
\bea\lb{fulsup}
\td W&=&\Tr\left(-\zeta\td\Phi+\td\Phi(\td YY-M_{xx})
+h (M_{xz}M_{zx}-M_{zz}Y\td Y+\zeta M_{zz})\right)\nonumber \\[3mm]
&+&\ld\left(
\det\left(
\ba{cc}
M_{xx}&M_{xz}\\
M_{zx}&M_{zz}
\ea
\right)
-AB-\Ld^{4N+2M}
\right)
\eea
Here $\lambda$ is a lagrange multiplier field. Its constraint equation is the quantum deformed
relation between the baryon and meson fields, and dictates that either the baryons or mesons
acquire a non-zero vev. We assume that we are on the baryonic branch
\be\lb{barb}
AB=-\Ld^{4N+2M}\, .
\ee
The vevs of the baryons break the non-anomalous $U(1)_B$.
The D-term equations for $U(1)_B$ fix $|A|^2=|B|^2$.

The adjoint field $\td\Phi$, and the meson fields $M_{xx}$, $M_{xz}$, and $M_{zx}$ are all massive.
So we can integrate them out.
The reduced gauge theory has gauge group $SU(N) \times SU(N+M)$, a pair of bi-fundamental
fields $(Y, \td Y)$, and a meson field
\be
\Phi = M_{zz}
\ee
that transforms as an adjoint under $SU(N+M)$.
After integrating out the massive fields, the superpotential
(\ref{fulsup}) reduces to the ISS superpotential in the magnetic
regime \cite{ISS}
\bea
\td W=h \, \zeta\,
\Tr( \Phi) - h\, \Tr\bigl( \Phi Y\td Y\bigr)\, ,
\eea
Up to relabeling the nodes 1 and 3, this quiver gauge theory coincides
with the quiver in figure \ref{spp-geom1}.

\end{document}